\documentclass[preprint,showpacs,preprintnumbers,amsmath,amssymb]{revtex4}

\usepackage{graphicx}
\usepackage{dcolumn}
\usepackage{bm}

\begin{document}


\title{Gauge Theory Model of the Neutrino  and \\ New Physics Beyond the Standard Model}
\author{Yue-Liang Wu }
\email{ylwu@itp.ac.cn}
\affiliation{ KITPC/ITP-CAS
\\
Kavli Institute for Theoretical Physics China,Institute of
Theoretical Physics \\ Chinese Academy of sciences, Beijing 100190,
China}
 %
\date{}
\begin{abstract}
Majorana features of neutrinos and SO(3) gauge symmetry of three
families enable us to construct a gauge model of neutrino for
understanding naturally the observed smallness of neutrino masses
and the nearly tri-bimaximal neutrino mixing when combining together
with the mechanism of approximate global U(1) family symmetry. The
vacuum structure of SO(3) symmetry breaking is found to play an
important role. The mixing angle $\theta_{13}$ and CP-violating
phases governed by the vacuum of spontaneous symmetry breaking are
in general non-zero and testable experimentally at the allowed
sensitivity. The model predicts the existence of vector-like SO(3)
triplet charged leptons and vector-like SO(3) triplet Majorana
neutrinos as well as SO(3) tri-triplet Higgs bosons, some of them
can be light and explored at the colliders LHC and ILC.
\end{abstract}

\maketitle

The evidence of massive neutrinos strongly indicates new physics
beyond the standard model\cite{MS}. The dominance of dark matter in
our universe challenges the standard models of both particle physics
and cosmology. The current neutrino experimental
data\cite{EXP1,EXP2,EXP3,EXP4,EXP5,EXP6,EXP7,EXP8,PDG} can well be
described by neutrino oscillations via three neutrino
mixings\cite{TNM1,TNM2,SV}. The global fit from various experimental
data lead to the following constraints on the three mixing angles:
\begin{eqnarray}
30\mbox{}^{\circ}<\theta_{\mathrm{12}}<38\mbox{}^{\circ}, \quad
36\mbox{}^{\circ}<\theta_{\mathrm{23}}<54\mbox{}^{\circ}, \quad
\theta_{\mathrm{13}}<10\mbox{}^{\circ} \nonumber
\end{eqnarray}
and the mass squared differences:
\begin{eqnarray}
& &  7.2\times{10}^{-5}\ {\mathrm{eV}}^{2}    <\Delta
m_{21}^{2}=m_{\nu_{\mu}}^2 - m_{\nu_e}^2 <8.9\times {10}^{-5}\ {\mathrm{eV}}^{2}, \nonumber \\
& & 2.1\times{10}^{-3}\ {\mathrm{eV}}^{2}    <\Delta m_{32}^{2}=
m_{\nu_{\tau}}^2 - m_{\nu_{\mu}}^2 <3.1\times {10}^{-3}\
{\mathrm{eV}}^{2} \nonumber
\end{eqnarray}
at the 99\% confidence level\cite{SV}.

The most unclear parameter is the mixing angle
$\theta_{\mathrm{13}}$ which is expected to be measured in near
future. Phenomenologically, such mixing angles are consistent with
the so-called tri-bimaximal mixing with $\theta_{12} =
\sin^{-1}(1/\sqrt{3}) = 35^{\circ}$, $\theta_{23} =
\sin^{-1}(1/\sqrt{2}) = 45^{\circ}$ and $\theta_{13} =0$, which was
first proposed by Harrison, Perkins and Scott \cite{HPS} and
investigated further by many groups\cite{HPS1,HPS2,HPS3,HPS4}. Great
theoretical efforts have been made to obtain such a mixing matrix
via imposing various
symmetries\cite{TBM1,TBM2,TBM3,TBM4,TBM5,TBM6,TBM7,TBM8,TBM9,TBM10,TBM11,TBM12,TBM13,TBM14,TBM15,TBM16,TBM17}.
However, for $\theta_{13} =0$, it is hard to be directly tested
experimentally. Recently, it was shown that the tri-bimaximal
neutrino mixing matrix may be yielded as the lowest order
approximation from diagonalizing a special symmetric mass matrix for
the Dirac-type neutrinos with a new symmetry\cite{FLee} and the
Majorana-type neutrinos with the $Z_3$ group\cite{HWW}, where the
mixing angle $\theta_{13}$ is in general non-zero and can be at the
experimentally allowed sensitivity. Nevertheless, in comparison with
the quark sector, one still faces a puzzle that why neutrino masses
are so tiny, but their mixings are so large. It is known that the
only peculiar property for neutrinos is that they could be Majorana
fermions. It is then natural to conjecture that a solution to the
puzzle is most likely attributed to the Majorana features. Thus
revealing the origin of large mixing angles and small masses of
neutrinos is important not only for understanding neutrino physics,
but also for exploring new physics beyond the standard model.

The greatest success of the standard model (SM) is the gauge
symmetry structure $SU(3)_{c}\times SU_{L}(2)\times U_{Y}(1)$ which
has been tested by more and more precise experiments. As a simple
extension of the standard model with three families and Majorana
neutrinos, we are going to consider in this note a non-abelian gauge
family symmetry SO(3) instead of discrete symmetries discussed
widely in literature. It is noted that only SO(3) rather than SU(3)
is allowed due to the Majorana feature of neutrinos. In fact, it was
shown that SO(3) family symmetry can easily explain the maximal
mixing between muon-neutrino and tau-neutrino, i.e., $\theta_{23} =
45^{\circ}$ and nearly degenerate neutrino
masses\cite{YLW1,YLW2,YLW3,CS,MA,CW,BHKR}, it can also lead to a
nearly bi-maximal neutrino mixing (based on the early data) relying
on the SO(3) triplet scalar fields and the symmetry breaking
scenarios of SO(3) symmetry\cite{YLW4,YLW5,YLW6}. As the current
experimental data favor a nearly tri-bimaximal mixing, we shall
construct in this note an alternative model with gauge symmetry
$SO(3)\times SU(3)_c\times SU_L(2)\times U_Y(1)$. The new model
contains a minimal set of new particles which include vector-like
SO(3) triplet Majorana fermion and vector-like SO(3) triplet charged
lepton, two $SU(2)_L$ Higgs doublets, two SO(3) tri-triplet Higgs
bosons and one singlet Higgs boson. By considering an appropriate
symmetry breaking scenario of SO(3) family symmetry, the nearly
tri-bimaximal neutrino mixing matrix is naturally obtained. It is
different from our previous consideration for obtaining the nearly
bimaximal neutrino mixing, where we introduced only the SO(3)
triplet Higgs bosons. In this note, we shall show that it is more
natural and simple to introduce SO(3) tri-triplet Higgs bosons
instead of SO(3) triplet Higgs bosons to obtain the nearly
tri-bimaximal neutrino mixing. In particular, the type II like
seesaw mechanism is easily realized in such a model, and the
smallness of neutrino masses and nearly tri-bimaximal neutrino
mixing can well be understood simultaneously via the symmetry
breaking scenario of SO(3) tri-triplet Higgs bosons and the
mechanism of approximate global U(1) family symmetry\cite{HW,WW,WU}
in the Yukawa sector. Some new particles can be light and probed in
future experiments.

For our purpose in this note, let us focus on the following
$SO(3)\times SU(2)_{L}\times U(1)_{Y}$ invariant Lagrangian for
Yukawa interactions of leptons with Majorana neutrinos
\begin{eqnarray}
{\cal L}_Y & = & y_{\nu} \bar{l} \tilde{H}\nu_R + y_{N} \bar{l} H_N
N  + \frac{1}{2}\xi_{N} \bar{N} \Phi_{\nu} N  +
\frac{1}{2} M_R \bar{\nu}_R\nu_R^c  \nonumber \\
& + & y_e \bar{l} H E + \xi_e \phi_s \bar{E} e_R +  \frac{1}{2}
\xi_E \bar{E} \Phi_e E + H.c.
\end{eqnarray}
with $y_{\nu}$, $y_e$, $y_N$, $\xi_e$, $\xi_N$ and $\xi_E$ being the
real Yukawa coupling constants. $M_R$ is the mass of right-handed
Majorana neutrinos. All the fermions $\nu_{Li}$, $\nu_{Ri}$,
$e_{Li}$, $e_{Ri}$, $E_i$ and $N_{i}$ $(i=1,2,3)$ belong to SO(3)
triplets in family space. Where $\bar{l}_i= (\bar{\nu}_{Li},
\bar{e}_{Li} )$ denote $SU_L(2)$ doublet leptons, $H$ and $H_N$ are
$SU_L(2)$ doublet Higgs bosons with $\tilde{H} = \tau_2 H^*$.
$\nu_{Ri}$ are the right-handed neutrinos with $\nu_{Ri}^c =
c\bar{\nu}_{Ri}^T$ the charge conjugated ones. $E_i$ are $SU_L(2)$
singlet vector-like charged leptons and the $N_{i}$ are $SU_L(2)$
singlet vector-like Majorana neutrinos with $N_{ i}^c = N_{i}$.
$\phi_s$ is a singlet Higgs boson. The scalar fields $\Phi_{\nu}$
and $\Phi_e$ are SO(3) tri-triplets Higgs bosons. The hermiticity
condition of Lagrangian and the Majorana condition of vector-like
neutrinos imply that
\begin{eqnarray}
\Phi_{\nu} = \Phi_{\nu}^{\ast},\quad \Phi_{\nu} = \Phi_{\nu}^T,
\quad \Phi_e = \Phi_e^{\dagger}
\end{eqnarray}
Namely $\Phi_{\nu}$ is a real symmetric tri-triplet Higgs boson and
contains six independent scalar fields, $\Phi_e$ is an Hermitian
tri-triplet Higgs boson and contains nine independent scalar fields.
With the above fields, the Lagrangian in eq.(1) is the most general
one ensured by the following discrete symmetry ($Z_2$ and $Z_4$)
\begin{eqnarray}
N \to i\gamma_5\ N, \quad \Phi_{\nu} \to - \Phi_{\nu}, \quad H_N \to
-iH_N, \quad \phi_s \to - \phi_s,\quad e_R \to - e_R
\end{eqnarray}

We now discuss the interesting features of SO(3) gauge symmetry. In
terms of SO(3) representation, one can reexpress the real symmetric
tri-triplet Higgs boson into the following general form
\begin{eqnarray}
& & \Phi_{\nu}  \equiv  O_{\nu} \phi_{\nu} O_{\nu}^T, \quad
O_{\nu}(x) = e^{i\lambda^i \Theta^{\nu}_i(x)},  \quad \phi_{\nu}(x)
= \left(
                     \begin{array}{ccc}
                       \phi_1^{\nu} & \phi_2^{\nu} & \phi_3^{\nu} \\
                       \phi_2^{\nu} & \phi_3^{\nu} & \phi_1^{\nu} \\
                       \phi_3^{\nu} & \phi_1^{\nu} & \phi_2^{\nu} \\
                     \end{array}
                   \right)
\end{eqnarray}
with $\lambda^i$ $(i=1,2,3)$ being the generators of SO(3). Where
$\Theta_i^{\nu}(x)$ $(i=1,2,3)$ may be regarded as three rotational
scalar fields of SO(3), and $\phi_i^{\nu}(x)$ $(i=1,2,3)$ are the
remaining three independent scalar fields.

The non-trivial structure of $\phi_{\nu}(x)$ is a unique property of
the cyclic Abelian finite group $Z_3$ for a real symmetric matrix
with three independent elements. A detailed discussion on the
properties of $Z_3$ group can be found in ref.\cite{HWW}. The
explicit three dimensional unitary representation of $Z_3=\{ t_i \}$
$(i=1,2,3)$ is given by
\begin{align*}
t_{1}  &  =\left(
\begin{array}
[c]{ccc}%
1 & 0 & 0\\
0 & 1 & 0\\
0 & 0 & 1
\end{array}
\right) ,\  t_{2}=\left(
\begin{array}
[c]{ccc}%
0 & 1 & 0\\
0 & 0 & 1\\
1 & 0 & 0
\end{array}
\right) ,\ t_{3}=\left(
\begin{array}
[c]{ccc}%
0 & 0 & 1\\
1 & 0 & 0\\
0 & 1 & 0
\end{array}
\right)  .
\end{align*}
which is the only nontrivial invariant subgroup of the non-Abelian
symmetric group $S_3=\{ t_i,\ T_i \}$ $(i=1,2,3)$ with
\begin{align*}
T_{1}  &  =\left(
\begin{array}
[c]{ccc}%
1 & 0 & 0\\
0 & 0 & 1\\
0 & 1 & 0
\end{array}
\right),\ T_{2}=\left(
\begin{array}
[c]{ccc}%
0 & 1 & 0\\
1 & 0 & 0\\
0 & 0 & 1
\end{array}
\right),\  T_{3}  =\left(
\begin{array}
[c]{ccc}%
0 & 0 & 1\\
0 & 1 & 0\\
1 & 0 & 0
\end{array}
\right).
\end{align*}
It is easy to check that the non-trivial structure of
$\phi_{\nu}(x)$ belongs to the coset space of the symmetric group
$S_3$ with $Z_3$ as the subgroup, i.e., $\phi_{\nu}(x) \in S_3/Z_3$.
To be more explicit, the symmetric scalar field $\phi_{\nu}(x)$ with
three independent components can be expressed in terms of the group
representation $\{ T_i |i=1,2,3\} \in S_3/Z_3$ as the follows
\begin{eqnarray}
\phi_{\nu}(x) = \phi_1^{\nu}(x) T_1 +   \phi_2^{\nu}(x) T_2 +
\phi_3^{\nu}(x) T_3
\end{eqnarray}
which is in a cyclic permuted form $[\phi_{\nu}(x)]_{ij} =
\phi^{\nu}_{i+j-1}(x)$ with $(i+j-1)$ mod. 3 and invariant under the
$Z_3$ operation
\[t \phi_{\nu} t = \phi_{\nu}\]

SO(3) gauge invariance allows us to fix the gauge by making SO(3)
gauge transformation $g(x)$ to satisfy the condition $g(x) \equiv
O_{\nu}(x) \in SO(3)$. With such a gauge fixing, the Yukawa
interactions can be rewritten as follows
\begin{eqnarray}
{\cal L}_Y & = & y_{\nu} \bar{l} \tilde{H}\nu_R + y_{N} \bar{l} H_N
N + \frac{1}{2}\xi_{N} \bar{N} \phi_{\nu} N  +
\frac{1}{2} M_R \bar{\nu}_R\nu_R^c   \nonumber \\
& + & y_e \bar{l} H E + \xi_e \phi_s \bar{E} e_R +  \frac{1}{2}
\xi_E \bar{E} \hat{\Phi}_e E + H.c.
\end{eqnarray}
where $\hat{\Phi}_e = O_{\nu}^T \Phi_e  O_{\nu}$ remains Hermitian
and contains nine independent scalar fields. Note that the resulting
Lagrangian with such a gauge fixing is invariant under $Z_3$
transformation.

The Hermitian SO(3) tri-triplet scalar field $\hat{\Phi}_e(x)$ can
generally be reexpressed in terms of SO(3) representation as the
following form
\begin{eqnarray}
 \hat{\Phi}_e \equiv U_e \phi_e U_e^{\dagger},
 \quad U_e(x) \equiv P_e O_{e}, \quad \quad O_e(x) =
e^{i\lambda^i \chi^{e}_i(x)},
\end{eqnarray}
and
\begin{eqnarray}
& & P_e(x)  =  \left(
 \begin{array}{ccc}
  e^{i\eta_1^e(x)} & 0 & 0 \\
    0 & e^{i\eta_2^e(x)} & 0 \\
    0 & 0 & e^{i\eta_3^e(x)}
  \end{array}
 \right),\quad \phi_{e}(x) = \left(
                     \begin{array}{ccc}
                       \phi_1^{e}(x) & 0 & 0 \\
                       0 & \phi_2^{e}(x) & 0 \\
                       0 & 0 & \phi_3^{e}(x) \\
                     \end{array}
                    \right)
\end{eqnarray}
where $\chi_i^e(x)$ $(i=1,2,3)$ are regarded as three rotational
scalar fields of SO(3), $\eta_i^e(x)$ $(i=1,2,3)$ denote three phase
scalar fields and $\phi_i^e(x)$ $(i=1,2,3)$ are the remaining three
independent scalar fields.

When all the scalar fields evaluate their vacuum expectation values,
both SO(3) and $SU_L(2)$ gauge symmetries and the discrete
symmetries are broken down spontaneously. Let us take the previously
given gauge fixing condition and consider the following general
vacuum structure of scalar fields
\begin{eqnarray}
& & <H(x)> = v, \qquad <H_N(x)> = v_N \\
& &  <\phi_s(x)> = v_s, \qquad <\phi_i^{\nu}(x)> = v_i^{\nu},  \\
& &  <\phi_i^e(x)> = v_i^e,\qquad <\chi_i^e(x)> = \theta_i^e , \quad
<\eta_i^e(x)> = \delta_i^e
\end{eqnarray}
namely
\begin{eqnarray}
& & <P_e> \equiv P_{\delta}^e  = diag.( e^{i\delta_1^e},
e^{i\delta_2^e}, e^{i\delta_3^e} ), \quad <O_e> = e^{i\lambda^i
\theta_i^e}
\end{eqnarray}
Where $\delta_i^e$ (i=1,2,3) are CP phases arising from spontaneous
symmetry breaking and $\theta_i^e$ are three rotational angles of
SO(3).

With such a vacuum structure after spontaneous symmetry breaking,
the mass matrix of neutrinos and charged leptons are given by the
following generalized see-saw mechanism
\begin{eqnarray}
& & M_{\nu} = m^D_{\nu} M_R^{-1} m^D_{\nu} + m^D_{N} M_N^{-1}
m^D_{N}, \\
& &  M_e =  V_e m_E^D M_E^{-1} m_E^D V_e^{\dagger}
\end{eqnarray}
with
\begin{eqnarray}
& & m^D_{\nu} = y_{\nu} v, \quad m^D_{N} = y_N v_N , \quad m_E^D = \sqrt{y_{e} \xi_e v  v_s } \\
& & V_e = <U_{e}> = P_{\delta}^e e^{i \lambda^i \theta_i^e }
\end{eqnarray}
and
\begin{eqnarray}
 & & M_N = \xi_{N} \left(
        \begin{array}{ccc}
          v_1^{\nu} &  v_2^{\nu} &  v_3^{\nu} \\
           v_2^{\nu} &  v_3^{\nu} &  v_1^{\nu} \\
           v_3^{\nu} &  v_1^{\nu} &  v_2^{\nu} \\
        \end{array}
      \right),\qquad  M_E = \xi_E \left(
            \begin{array}{ccc}
               v_1^{e} & 0 & 0 \\
              0 &  v_2^{e} & 0 \\
              0 & 0 &  v_3^{e} \\
            \end{array}
          \right) \\
& & V_e \equiv
               P_{\delta}^e \left(
                     \begin{array}{ccc}
                       c_{12}^ec_{13}^e\ \ & s_{12}^ec_{13}^e\ \ & s_{13}^e \\
                       -s_{12}^ec_{23}^e-c_{12}^es_{23}^es_{13}^e\  \ &
                       c_{12}^ec_{23}^e - s_{12}^es_{23}^es_{13}^e\ \ & s_{23}^e c_{13}^e \\
                       s_{12}^e s_{23}^e - c_{12}^ec_{23}^es_{13}^e\ \ & -c_{12}^es_{23}^e - s_{12}^ec_{23}^es_{13}^e\  \
                       & c_{23}^ec_{13}^e \\
                     \end{array}
                   \right)
\end{eqnarray}
Where we have used the notations $c_{ij}^e \equiv \cos\theta_{ij}^e$
and $s_{ij}^e \equiv \sin\theta_{ij}^e$. Note that $\theta_{ij}^e$
are given as functions of $\theta_i^e$ $(i=1,2,3)$.

It is of interest to notice that when taking the Majorana neutrino
masses $M_R$ and $M_N$ to be infinity large, the interactions with
Majorana neutrinos decouple from the theory. This can be seen from
the following effective interactions mediated via the Majorana
neutrinos
\begin{eqnarray}
\frac{y^2_{\nu}}{M_R} \bar{l}\tilde{H} \tilde{H}^{T} l^c, \ \
\frac{y^2_{N}}{M_N} \bar{l}H_N H_N^{T} l^c \to 0, \quad \mbox{for}
\quad M_R,\   \  M_N \to \infty
\end{eqnarray}
which implies that the resulting Yukawa interactions in this limit
generate additional global U(1) family symmetries for the charged
lepton sector. Namely, once the Majorana neutrinos become very
heavy, the Yukawa interactions possess approximate global U(1)
family symmetries.

When applying the mechanism of approximate global U(1) family
symmetries\cite{HW,WW,WU} to the Yukawa interactions after SO(3)
symmetry is broken down spontaneously, we arrive at the following
conditions
\begin{eqnarray}
& & M_R \gg m^D_{\nu},\qquad  M_N \gg m_N^D, \qquad \theta_i^e \ll
1,
\end{eqnarray}
Namely
\begin{eqnarray}
 M_{\nu} \ll 1, \qquad \theta_{ij}^e \ll 1
\end{eqnarray}
which provides a possible explanation why the observed left-handed
neutrinos are so light and meanwhile the charged lepton mixing
angles must be small. With the above conditions, the neutrino mass
matrix (eq.(13)) is given by a type II like see-saw mechanism and
the charged lepton mass matrix (eq.(14)) is also presented by a
generalized see-saw mechanism. By diagonalizing the mass matrices,
we have
\begin{eqnarray}
V_{\nu}^{T} M_{\nu} V_{\nu} = diag.(m_{\nu_e}, m_{\nu_{\mu}},
m_{\nu_{\tau}} ), \quad V_e^{\dagger} M_e V_e = diag.(m_e, m_{\mu},
m_{\tau} )
\end{eqnarray}
where
\begin{eqnarray}
V_{\nu} = \left(
  \begin{array}{ccc}
    \frac{2}{\sqrt{6}}c_{\nu}\  \
     & \frac{1}{\sqrt{3}}\  \  & \frac{2}{\sqrt{6}}s_{\nu} \\
    -\frac{1}{\sqrt{6}}c_{\nu} - \frac{1}{\sqrt{2}}s_{\nu}\  \  & \frac{1}{\sqrt{3}}\  \  &
    \frac{1}{\sqrt{2}}c_{\nu} - \frac{1}{\sqrt{6}}s_{\nu} \\
    -\frac{1}{\sqrt{6}}c_{\nu} + \frac{1}{\sqrt{2}}s_{\nu}\  \  & \frac{1}{\sqrt{3}}\  \  &
    - \frac{1}{\sqrt{2}}c_{\nu}-\frac{1}{\sqrt{6}}s_{\nu}  \\
  \end{array}
\right) \equiv V_0 V_1
\end{eqnarray}
with
\begin{eqnarray}
 V_0 = \left(
  \begin{array}{ccc}
    \frac{2}{\sqrt{6}}
     & \frac{1}{\sqrt{3}} & 0 \\
    -\frac{1}{\sqrt{6}} & \frac{1}{\sqrt{3}} & \frac{1}{\sqrt{2}} \\
    -\frac{1}{\sqrt{6}} & \frac{1}{\sqrt{3}} & -\frac{1}{\sqrt{2}} \\
  \end{array}
\right), \quad V_1 = \left(
        \begin{array}{ccc}
          c_{\nu} & 0 & s_{\nu} \\
          0 & 1 & 0 \\
          -s_{\nu} & 0 & c_{\nu} \\
        \end{array}
      \right)
\end{eqnarray}
Here $V_0$ is the so-called tri-bimaximal mixing matrix\cite{HPS}.
For short, we have introduced the notations $c_{\nu} \equiv \cos
\theta_{\nu}$ and $s_{\nu} \equiv \sin \theta_{\nu}$ with
\begin{eqnarray}
\tan 2 \theta_{\nu} = \frac{\sqrt{3}(v_{21}^{\nu} - v_{31}^{\nu})
}{v_{21}^{\nu} + v_{31}^{\nu}}, \quad v_{21}^{\nu} \equiv
v_{2}^{\nu} - v_{1}^{\nu}, \quad v_{31}^{\nu} \equiv v_{3}^{\nu} -
v_{1}^{\nu}
\end{eqnarray}
As a consequence, the leptonic CKM-type mixing matrix in the mass
eigenstates of leptons is given by
\begin{eqnarray}
V & = & V_e^{\dagger} V_{\nu}  =
                       \left(
                       \begin{array}{ccc}
                       c_{12}^ec_{13}^e\ \ & -s_{12}^ec_{23}^e-c_{12}^es_{23}^es_{13}^e\  \
                        & s_{12}^e s_{23}^e - c_{12}^ec_{23}^es_{13}^e \\
                       s_{12}^ec_{13}^e \  \ & c_{12}^ec_{23}^e - s_{12}^es_{23}^es_{13}^e  \ \
                       & -c_{12}^es_{23}^e - s_{12}^ec_{23}^es_{13}^e \\
                       s_{13}^e \ \ & s_{23}^e c_{13}^e  \  \
                       & c_{23}^ec_{13}^e \\
                     \end{array}
                   \right)  \nonumber \\
 & & \qquad  \times \left(
             \begin{array}{ccc}
               e^{-i\delta_1^e} & 0 & 0 \\
               0 &  e^{-i\delta_2^e}  & 0 \\
               0 & 0 &  e^{-i\delta_3^e}  \\
             \end{array}
           \right)
            \left(
  \begin{array}{ccc}
    \frac{2}{\sqrt{6}}
     & \frac{1}{\sqrt{3}} & 0 \\
    -\frac{1}{\sqrt{6}} & \frac{1}{\sqrt{3}} & \frac{1}{\sqrt{2}} \\
    -\frac{1}{\sqrt{6}} & \frac{1}{\sqrt{3}} & -\frac{1}{\sqrt{2}} \\
  \end{array}
\right) \left(
        \begin{array}{ccc}
          c_{\nu} & 0 & s_{\nu} \\
          0 & 1 & 0 \\
          -s_{\nu} & 0 & c_{\nu} \\
        \end{array}
      \right) \\
      & \equiv &   P_{\beta}^e \left(
                     \begin{array}{ccc}
                       c_{12}c_{13}\ \ & s_{12}c_{13}\ \ & s_{13}e^{-i\delta} \\
                       -s_{12}c_{23}-c_{12}s_{23}s_{13}e^{i\delta}\  \ &
                       c_{12}c_{23} - s_{12}s_{23}s_{13}e^{i\delta} \ \ & s_{23} c_{13} \\
                       s_{12} s_{23} - c_{12}c_{23}s_{13}e^{i\delta} \ \
                       & -c_{12}s_{23} - s_{12}c_{23}s_{13} e^{i\delta}\  \
                       & c_{23}c_{13} \\
                     \end{array}
                   \right) \left(
             \begin{array}{ccc}
                e^{i\alpha_1} & 0 & 0 \\
               0 &  e^{i\alpha_2} & 0 \\
               0 & 0 &  1 \\
             \end{array}
           \right)
\end{eqnarray}
with $P_{\beta}^e = diag.\left( e^{i\beta_1^e}, e^{i\beta_2^e},\
e^{i\beta_3^e} \right)$, and $c_{ij}\equiv \cos\theta_{ij}$ and
$s_{ij}\equiv \sin\theta_{ij}$. Where $\beta_i^e$ (i=1,2,3) are
three phases introduced to parameterize the leptonic mixing matrix
into a familiar form which has been used widely for quark mixing
matrix. The phases $\alpha_1$ and $\alpha_2$ are known as Majorana
phases. One can relate three mixing angles $\theta_{ij}$ and CP
phases $\beta_i^e$, $\delta$, $\alpha_1$ and $\alpha_2$ with mixing
angles $\theta_{ij}^e$, $\theta_{\nu}$, phases $\delta_{i}^e$. Three
phases $\beta_i^e$ can in principle be absorbed by the phase
redefinitions of charged leptons. Nevertheless, unlike in the
standard model, the phases $\beta_i^e$ cannot be rotated away in the
model due to SO(3) gauge interactions among three families, their
physical effects will occur in processes involving SO(3) gauge
interactions.

It is seen that the smallness of $\theta_{ij}^e$ $(i<j)$ is
attributed to the mechanism of approximate global U(1) family
symmetries in the Yukawa sector. Thus in a good approximation up to
the first order of $s_{ij}^e$, the leptonic mixing matrix may be
expressed as the following simplified form
\begin{eqnarray}
V & \simeq & \left(
                       \begin{array}{ccc}
                       1 \ & -s_{12}^e \
                        & - s_{13}^e \\
                       s_{12}^e \ &  1 \
                       & -s_{23}^e  \\
                       s_{13}^e \ &  s_{23}^e  \
                       & 1 \\
                     \end{array}
                   \right) P_{\delta}^{e \dagger}
           \left(
  \begin{array}{ccc}
    \frac{2}{\sqrt{6}}
     & \frac{1}{\sqrt{3}} & 0 \\
    -\frac{1}{\sqrt{6}} & \frac{1}{\sqrt{3}} & \frac{1}{\sqrt{2}} \\
    -\frac{1}{\sqrt{6}} & \frac{1}{\sqrt{3}} & -\frac{1}{\sqrt{2}} \\
  \end{array}
\right) \left(
        \begin{array}{ccc}
          c_{\nu} & 0 & s_{\nu} \\
          0 & 1 & 0 \\
          -s_{\nu} & 0 & c_{\nu} \\
        \end{array}
      \right)
\end{eqnarray}

The three vector-like heavy Majorana neutrino masses are obtained
via diagonalizing the mass matrix $M_N$, i.e., $V^{\dagger}_{\nu}
M_N V_{\nu} = diag.(m_{N_1}, \ m_{N_2},\ m_{N_3})$
\begin{eqnarray}
m_{N_1} = - m_N \sqrt{1-\Delta},\quad m_{N_2} = m_N, \quad m_{N_3} =
m_N \sqrt{1-\Delta}
\end{eqnarray}
with
\begin{eqnarray}
& & m_N \equiv \xi_N (v_1^{\nu} + v_2^{\nu} + v_3^{\nu} ) \\
& & \Delta = 3(v_1^{\nu}v_2^{\nu} + v_2^{\nu}v_3^{\nu} +
v_3^{\nu}v_1^{\nu})/(v_1^{\nu} + v_2^{\nu} + v_3^{\nu})^2,
\end{eqnarray}
Note that the minus sign in $m_{N_1}$ occurs when $(v_2 + v_3) > 2
v_1$ and can be absorbed by the redefinition of the Majorana
neutrino $N_{1} \to i\gamma_5 N_{1}$. If $(v_2 + v_3) < 2 v_1$, the
mass of $N_3$ becomes negative. It is noticed that the Majorana
neutrinos $N_1$ and $N_3$ are in general degenerate in their masses
in the physical basis.

The masses of three left-handed light Majorana neutrinos are given
in the physics basis as follows
\begin{eqnarray}
& & m_{\nu_e} = \bar{m}_{0} - m_1 ( 2 + \bar{\Delta})  \\
& & m_{\nu_{\mu}} = \bar{m}_{0}  \\
& & m_{\nu_{\tau}} = \bar{m}_{0} + m_1 \bar{\Delta}
\end{eqnarray}
with
\begin{eqnarray}
& & \bar{m}_{0} \equiv m_0 + m_{1},\qquad \bar{\Delta} =
1/\sqrt{1-\Delta} - 1 \\
& & m_0 = (m^D_{\nu})^2/M_R, \qquad m_1 = (m_N^D)^2/m_N,
\end{eqnarray}

With the above analysis, it enables us from the experimentally
measured neutrino mass squire differences to extract two mass
parameters $m_0$ and $m_1$ for a given value of parameter $\Delta$
with $\Delta \neq 1$\footnote{Note that when all the vacuum
expectation values are equal, i.e., $v_1^{\nu} = v_2^{\nu} =
v_3^{\nu}$, the vector-like Majorana neutrino mass matrix becomes
democratic and $\Delta = 1$. As a consequence, two of the
vector-like Majorana neutrinos become massless in the physical basis
and the type II like see-saw mechanism cannot be realized to obtain
the observed neutrino masses and mixing.}. To be explicit, in terms
of the parameter $\bar{\Delta}$ and the neutrino mass squire
differences, we obtain the following relations
\begin{eqnarray}
& & m_1 = \sqrt{(m_{\nu_{\tau}}^2 - m_{\nu_{\mu}}^2)/\left((2R_1 +
\bar{\Delta})\bar{\Delta}\right) },\qquad m_0 = (R_1 -1)  m_1
\end{eqnarray}
with
\begin{eqnarray}
 & & R_1 = \left(1 +
\frac{\bar{\Delta}}{2} + \frac{\bar{\Delta}^2}{2(2+ \bar{\Delta})}\
R_0 \right)/\left( 1 - \frac{\bar{\Delta}}{2+ \bar{\Delta}}\ R_0
\right), \nonumber \\
& &  R_0 = (m_{\nu_{\mu}}^2 - m_{\nu_{e}}^2) / (m_{\nu_{\tau}}^2 -
m_{\nu_{\mu}}^2)
\end{eqnarray}

For a numerical calculation, we now present the light neutrino
masses as functions of the parameter $\Delta$ by taking the central
values of the mass squire differences $\Delta m_{21}^2 =
m_{\nu_{\mu}}^2 - m_{\nu_{e}}^2 = 8 \times 10^{-5}\ eV^2$ and
$\Delta m_{32}^2 = m_{\nu_{\tau}}^2 - m_{\nu_{\mu}}^2 = 2.5 \times
10^{-3}\ eV^2$ (see table 1).
\\
\\
Table 1: The masses of neutrino mass eigenstates $\nu_e$,
$\nu_{\mu}$, $\nu_{\tau}$ and mass parameters $m_0$ and $m_1$ as the
functions of the vacuum parameter $\Delta$
\\
\\
\begin{tabular}{|c|c|c|c|c|c|}
  \hline
  $\Delta$ (input) & $m_0 (10^{-2} eV)$ & $m_1 (10^{-2} eV)$ & $m_{\nu_e} ( 10^{-2} eV)$ & $m_{\nu_{\mu}}(10^{-2} eV)$
  & $m_{\nu_{\tau}}(10^{-2}eV)$ \\
  \hline
  0.75 & 1.297 & 2.487 & -3.677 & 3.784 & 6.271 \\
  0.73 & 1.271 & 2.637 & -3.908  & 4.003 & 6.346 \\
  0.71 & 1.245 & 2.789 & -4.034  & 4.333 & 6.424 \\
  0.69 & 1.220 & 2.943 & -4.163  & 4.666 & 6.506 \\
  \hline
\end{tabular}
\\
\\
where the minus sign of the Majorana neutrino mass can be absorbed
by a redefinition of the neutrino field $\nu_{e} \to i \nu_{e}$.

From the definition of the parameter $\Delta$ (eq.(31)), it is
determined by two ratios $v_1^{\nu}/v_2^{\nu}$ and
$v_3^{\nu}/v_2^{\nu}$ of three vacuum expectation values $v_i^{\nu}$
$(i=1,2,3)$. Also the mixing angle $\theta_{\nu}$ given in eq.(25)
is determined by the same two ratios. To relate the mixing angle
$\theta_{\nu}$ with the parameter $\Delta$ under the condition
$v_2^{\nu} + v_3^{\nu} > 2 v_1^{\nu}$, it is reasonable to consider
the interesting case with $v_1^{\nu} \ll v_2^{\nu}, v_3^{\nu}$. In
this case, the ratio $v_3^{\nu}/v_2^{\nu}$ becomes dominant and the
parameter $\Delta$ gets the maximal value $\Delta \leq 3/4$ at
$v_3^{\nu}=v_2^{\nu}$. Explicitly, we have the following approximate
relations when $v_1^{\nu} \ll v_2^{\nu}, v_3^{\nu}$
\begin{eqnarray}
& & \Delta \simeq \frac{3r}{(1+r)^2},\quad \tan 2 \theta_{\nu}
\simeq \frac{\sqrt{3}(1-r)}{1+r}, \quad r \equiv v_3^{\nu}/v_2^{\nu}
\end{eqnarray}
which shows that in this case both the mixing angle $\theta_{\nu}$
and the ratio $r$ can be determined for the given values of
$\Delta$. In the special case that $\Delta = r$, we arrive at the
solution
\begin{eqnarray}
\Delta = r = \sqrt{3} - 1, \quad \tan 2\theta_{\nu} = 2-\sqrt{3},
\quad \theta_{\nu} = 7.5^{\circ}
\end{eqnarray}

Let us now consider the leptonic mixing matrix element $V_{13}$
which has the following general form
\begin{eqnarray}
V_{13} \simeq \frac{2}{\sqrt{6}} s_{\nu}e^{-i\delta_1^e} +
\frac{1}{\sqrt{2}} c_{\nu} (s_{13}^e e^{-i\delta_3^e} - s_{12}^e
e^{-i\delta_2^e}) \equiv s_{13} e^{-i\delta_1^e -i\delta_{\nu}}
\end{eqnarray}
Here the definition of phase $\delta_{\nu}$ is, by convention, more
useful for discussing CP violation in neutrino sector. It is seen
that to completely determine the mixing angle $\theta_{13}$ and the
CP-violating phase $\delta_{\nu}$, one needs to know the charged
lepton mixing angles $s_{12}^e$ and $s_{13}^e$, the relative CP
phases $(\delta_1^e - \delta_2^e)$ and $(\delta_1^e - \delta_3^e )$.
For a numerical estimation, it is useful to investigate the
following two interesting cases
\begin{eqnarray}
& & \mbox{Case I}:\qquad s_{13}^e\simeq 0,\ s_{12}^e\simeq 0 \\
& & \mbox{Case II}:\qquad s_{13}^e \ll s_{12}^e \sim
\sqrt{m_e/m_{\mu}} \simeq 0.07, \quad \delta_1^e - \delta_2^e =\pi/2
\end{eqnarray}
Here the Case I ignores the charged lepton mixing angles for the
simplicity of analysis. The Case II is a typical case for the small
charged lepton mixing with assuming a maximal CP-violating phase
$(\delta_1^e - \delta_2^e) =\pi/2$. Though we are not able to
precisely predict the mixing angle $\theta_{13}$ and the
CP-violating phase $\delta_{\nu}$, while from the two typical cases,
it allows us to present a reasonable estimation for $\theta_{13}$
and $\delta_{\nu}$ (see table 2).
\\
\\
Table 2
\\
\\
\begin{tabular}{|c|c|c|c|c|c|c|}
  \hline
  $\Delta$ (input) & $r=v_3^{\nu}/v_2^{\nu}$ & $\theta_{\nu}$  & $\ \theta_{13}$ (Case I) & $\ \delta_{\nu}$ (Case I)
   & $\ \theta_{13}$ (Case II) & $\ \delta_{\nu}$ (Case II) \\
  \hline
  0.75 & 1.0 & 0 & 0 & 0 & $2.8^{\circ}$ & $90^{\circ}$ \\
  0.748 & 0.90 & $2.6^{\circ}$ & $2.1^{\circ}$ & 0 & $3.5^{\circ}$ & $54^{\circ}$ \\
  0.745 & 0.85 & $4.0^{\circ}$ & $3.3^{\circ}$ & 0 & $4.4^{\circ}$ & $41^{\circ}$ \\
  0.74 & 0.79 & $5.8^{\circ}$ & $4.7^{\circ}$ & 0 & $5.5^{\circ}$ & $31^{\circ}$ \\
  0.73 & 0.72 & $7.9^{\circ}$ & $6.4^{\circ}$ & 0 & $7.0^{\circ}$ & $24^{\circ}$ \\
  0.71 & 0.63 & $10.7^{\circ}$ & $8.7^{\circ}$ & 0 & $9.2^{\circ}$ & $18^{\circ}$ \\
  0.69 & 0.56 & $13.0^{\circ}$ & $10.6^{\circ}$ & 0 & $11.0^{\circ}$ & $15^{\circ}$ \\
  \hline
\end{tabular}
\\
\\
which indicates that as long as $v_3^{\nu}\neq v_2^{\nu}$ (i.e.,
$r\neq 1$) which should be a more general and reasonable case when
no symmetry is imposed, the mixing angle $\theta_{13}$ can be large
enough to be detected. For the case II, both mixing angle
$\theta_{13}$ and CP-violating phase $\delta_{\nu}$ are in general
testable by the future neutrino experiments.

From the above analysis, it is seen that as a reasonable
consideration that $v_1^{\nu} \ll v_2^{\nu},\ v_3^{\nu}$ with
$v_3^{\nu}\neq v_2^{\nu}$, one only needs to take the parameter
$\Delta$ to be slightly away from its maximal value $\Delta = 0.75$.
Numerically, taking $\Delta \simeq 0.72\pm 0.028$, we then arrive at
the most optimistic predictions for the mixing angle $\theta_{13}$
and CP-violating phase $\delta_{\nu}$
\begin{eqnarray}
& & \theta_{13}\simeq 7^{\circ}\pm 4^{\circ},\qquad \delta_{\nu}
\simeq 35^{\circ} \pm 20^{\circ}, \quad  r=v_3^{\nu}/
v_2^{\nu}\simeq 0.73\pm 0.17
\end{eqnarray}
which can be tested in the future experiments. It is seen that the
resulting value for the ratio $r=v_3^{\nu}/ v_2^{\nu}$ is also
reasonable.

We now turn to discuss the possible mass scales of new physics
beyond the standard model. When taking the Dirac neutrino masses
$m_{\nu}^D$ to be at the order of $(0.1\sim 1.0)$ MeV (i.e., at the
same order of electron mass), the mass scale $M_R$ of right-handed
heavy Majorana neutrino $\nu_R$ is found, from the values of $m_0$
in table 1, to be
\begin{eqnarray}
& & M_R \simeq O(1\sim 100)\ \mbox{TeV}
\end{eqnarray}
Similarly, if taking the Dirac type neutrino mass $m_N^D$ to be at
the same order $m_N^D\simeq (0.1\sim 1.0)$ MeV, the vector-like
Majorana neutrino masses are resulted from the values of $m_1$ in
table 1 to be
\begin{eqnarray}
& & m_{N_2} \equiv m_{N} \simeq 0.5 M_R \simeq O(500)\ \mbox{GeV} \sim O(50)\ \mbox{TeV}, \\
& & m_{N_1} = m_{N_3} = (0.50\sim 0.56) m_{N_2} \simeq O(250)\
\mbox{GeV} \sim O(25)\ \mbox{TeV}
\end{eqnarray}
which shows that the vector-like Majorana neutrino masses can be at
the electroweak scale in this case. In general, their masses can be
expected to be much smaller if the discrete symmetry imposed in
eq.(3) is only softly broken.

The masses for the vector-like charged leptons and their relations
to the SM charged leptons are given as follows
\begin{eqnarray}
& &  m_{E_i} = \xi_E v_i^e, \quad (i = 1,2,3) \\
& & m_{E_1} = (m_E^D)^2/ m_e , \quad  m_{E_2}= (m_E^D)^2/m_{\mu} ,
\quad m_{E_3} = (m_E^D)^2/m_{\tau}
\end{eqnarray}
which shows that the vector-like charged lepton masses have opposite
hierarchy to the SM charged lepton masses. It is seen that the heavy
triplet charged lepton masses depend on the mass parameter $m_E^D$.
For a numerical estimation, taking $m_E^D \simeq (15 \sim 25)$\ GeV,
we obtain the lightest vector-like charged lepton mass $m_{E_3}$ to
be
\begin{eqnarray}
& & m_{E_3} \simeq (127 \sim 352)\ \mbox{GeV}
\end{eqnarray}
which can be explored at LHC and ILC. The other two vector-like
charged leptons become very heavy with masses above TeV scale.

In conclusion, we have shown that the nearly tri-bimaximal neutrino
mixings and the smallness of neutrino masses can simultaneously be
understood in a flavor dynamical model with SO(3) gauge family
symmetry. It has been seen that the vacuum structure of SO(3)
symmetry breaking for the SO(3) tri-triplet Higgs bosons and the
mechanism of approximate global U(1) family symmetry play an
important role. The type II like seesaw mechanism is realized in
such a model. The mixing angle $\theta_{13}$ is in general nonzero
and its values can range from the experimentally allowed sensitivity
to the current experimental bound. CP violation in the lepton sector
is caused by a spontaneous symmetry breaking and can be
significantly large. Both the mixing angle $\theta_{13}$ and
CP-violating phase $\delta_{\nu}$ are expected to be testable by the
future more precise neutrino experiments. Some of the predicted
vector-like charged leptons and vector-like Majorana neutrinos as
well as new Higgs bosons in this model can be explored at the future
experiments as their masses can approach to the current experimental
bounds. In particular, the mechanism discussed in this note can be
extended to the quark sector for understanding the smallness of
quark mixing angles, which will be investigated elsewhere.

\acknowledgments

\label{ACK}

This work was supported in part by the National Science Foundation
of China (NSFC) under the grant 10475105, 10491306, and the key
Project of Chinese Academy of Sciences (CAS).


\end{document}